\documentclass{aastex}
\usepackage{spr-astr-addons}
\usepackage{url}

\RequirePackage{color}
\begin{document}
%
\title{Radiation Spectrum of a Magnetized Supercritical Accretion Disc with Thermal Conduction}


\author{M. Ghasemnezhad \altaffilmark{1}}
\and
\author{M. Khajavi\altaffilmark{1}}
\affil{}
\and
\author{S. Abbassi \altaffilmark{1,2}}
\affil{abbassi@ipm.ir}

\altaffiltext{1}{Department of Physics, School of Sciences, Ferdowsi University of Mashhad, Mashhad, 91775-1436, Iran}
\altaffiltext{2}{School of Astronomy, Institute for Research in Fundamental Sciences (IPM), Tehran, Iran}

\begin{abstract}
we examine the effect of thermal conduction on the observational properties of a super critical hot magnetized flow.
We obtained self-similar solution of a magnetized disc when the thermal conduction plays an important role.
Follow of our first paper (Ghasemnezhad et al. 2012 (hereafter GKA12)) we have extended our
solution on the observational appearance of the disc to
show how physical condition such as thermal conduction, viscosity, and advection will change the observed luminosity of the
disc, Continuous spectra and surface temperature of
such discs was plotted. We apply the present model to black-hole X-ray binary
LMC X-3 and narrow-line seyfert 1 galaxies, which are supposed
to be under critical accretion rate. Our results show clearly that the
surface temperature is strongly depends on the thermal
conduction, the magnetic field and advection parameter. However we
see that thermal conduction acts to oppose the temperature
gradient as we expect and observed luminosity of the disc will reduce when thermal conduction is high.
We have shown that in this model the spectra of critical accretion flows strongly depends on the inclination angle.
\end{abstract}


\section{INTRODUCTION}
Accretion discs are found around several kind of astrophysical
objects: from young stars (protoplanetary discs) to supermassive
black holes in AGNs. The dynamics of these discs is however
poorly understood. Accretion process is penetrated itself in various ways depending on that how
angular momentum of the discs it carries away and how the its energy is dissipated. Such systems typically under the effect of several physical process. It is known that, on average, matter moves inward resulting in the accretion of material onto the central object and powers the most luminous objects in the universe by converting gravitational potential energy into highly energetic radiation. Accretion disc forms when gaseous matter, usually an assembly of free electrons and various types of ions, spirals onto a central gravitating body by gradually losing its initial angular momentum effectively in a timescales compatible with observations, as a result of viscous and magnetic stresses. The energy released by accretion discs depends on the mass accretion rate as well as spin and mass of central black hole. There is a maximum value for luminosity (Eddington luminosity) of accretion disc which gravity is able to exceed the outward pressure of radiation.

The theory of black hole accretion disc has been developed
rapidly since 1970s and brings a lot of successes in describing
high-energy astronomical phenomena (See for a review Kato et al.
2008). As for thermally stable black hole accretion disc models,
there are only three classes have been developed. The fist one is
the standard Shakura $\&$ Sunyaev disc, which is optically thick
and geometrically thin, Keplerian rotation with very small infall
velocity, and truncated at the marginally stable orbit. The
standard disc was studied for the Newtonian case (e.g., Shakura
\& Sunyaev 1973) and for the relativistic case (e.g., Novikov \&
Thorne 1973). The second one is the optically thin ADAFs
(advection dominated accretion flows) that works very well for
small mass-accretion rates (Narayan \& Yi 1994) while for
supercritical accretion discs ($\dot{M}\sim\dot{M}_{crit}$) slim
discs was introduces (Abramowicz et al. 1988). $\dot{M}_{crit}$
is the critical mass accretion rate, defined by
$\dot{M}_{crit}=\frac{L_{E}}{c^{2}}$. $L_{E}$ and $c$ are the
Eddington luminosity and light speed, respectively. Therefore
this disc brightness about Eddington or super-Eddington
luminosity and the geometrical thickness cannot ignore
($\frac{H}{r}\leq 1$) (Abramowicz et al. 2010). The Slim
disc models describes supercritical disc accretion flow, so they
are identical to each other. Recently the slim disc model began
to attract considerable attention, because of its success to
reproduce spectra of accreting binaries. the spectral behaviour
of several black hole binaries cannot be explained by the
traditional standard disc, but is well reproduced by slim disc
models (Kubota 2001).  LMC X-3 is black hole (BH) binary system
in the large magellanic cloud (LMC) at a distance of 48.1 kpc
(derived from Orosz et al. 2009). Cowley et al. (1983) stablished
LMC X-3 as a BH condidate with a BH mass $7 M_{\odot}\leq M_{BH}
\leq 14 M_{\odot}$. LMC X-1 has a soft spectra and a low
absorption column density along the line of sight. These
properties make LMC X-3 an ideal sample for testing our
understanding of black hole accretion disc physics. Slim disk
model describes accretion flows at high luminosities, while
reducing to the standard thin disk in the low luminusity limit.
The most important feature of super critical accretion flows is
the presence of photon trapping as the advection of radiation
entropy. The Slim disc will exceed Eddington limit,
 because the accretion rate can be  larger than
Eddington accretion rate (Mineshige et al. 2000, Ohsuga et al.
2002). The basic stucture and spectral properties of the slim
disks have been examined by several researchers: Watarai \&
Fukue (1999) showed the existence of radiative winds from the
superdisk, Mineshige et al (2000) applied slim disks to the
narrow-line Seyfert 1 galaxies. Fukue (2000) examined the basic
properties of super critical accretion discs and derived the
relations between the observables and the model parameters.

Several physical process should be take into account when we are
studying black hole accretion disc. Magnetic field palys a great
role on the structure of the disc since in hot accreting flow,
the temperature is so high that the accreting materials are
ionized. Many researcher have been attack the problem to solve
the magnetohydrodynamics (MHD) equations of hot magnetized flow
analytically (Akizuki \& Fukue 2006, Abbassi et al. 2008,
Ghanbari et al 2007, Shadmehri \& khajenabi 2005). Kaburaki
(2000) has presented a set of analytical solutions for a hot
accretion flow in a global magnetic field.  Ghanbari et al.
(2007) have presented a set of self-similar solutions for
two-dimensional (2D) viscous-resistive ADAFs in the
presence of a dipolar magnetic field of the central accretor.
They have shown that the presence of a magnetic field and its
associated resistivity can considerably change the picture with
regard to accretion flows.

One of the largely neglected physical process in the physics of
accretion discs is thermal conduction; Hot accretion flows have
high temperature, so the internal energy per particle is high.
This is one of the reasons why advection cooling overcomes
radiative cooling. For the same reason turbulence heat transport
by thermal conduction is non-negligible in heat balance in the
disc (Kato et al. 2008).  while some recent observations
(Loewenstein et al. 2001; Di Matteo et al. 2003; Ho et al. 2003)
of the hot accretion flow around active galactic nuclei indicated
that it should be on collision-less regime. So thermal conduction
has a great role in energy transport in the accreting materials
in a hot accretion disc where they are completely ionized. The
weekly-collisional nature of hot acceretion flows has been noted
before (Mahadevan $ \& $ Quataret 1997). Since thermal conduction
act to oppose the formation of the temperature gradient that
causes it, that might expect that the temperature profile in a
thermal conducting disc should be differ from the case of the
disc was not under the influence of thermal conduction. Shadmehri
(2008), Abbassi et al. (2008, 2010), Tanaka $\& $ Menou (2006)
have studied the effect of hot accretion flow with thermal
conduction in a semi-analytical method; physics of such systems
have been studied in simulation models (e.g. Sharma et al. 2008;
Wu et al. 2010). Abbassi et al. (2008) have shown that for this
problem there are two types of solutions ; high and low accretion
rate. Ghasemnezhad et al. (2012) have shown, by putting an extra
physical condition have shown that the high accretion rate
solutions of Abbassi et al. (2008) are not exactly correct as
long as some of the low accretion rate solutions are not physical
meaning. This extra condition make a acceptable parameter space
which in this space the self-similar solutions has reasonable
physical behaviour. Following GKA12 we are investigating the
observational appearance a hot magnetized flow under effect of
thermal conduction using Fukue (2004) framework in a parameter
space presented by GKA12. We developed Fukue (2004) solutions for
narrow-line seyfert 1 galaxies by adding the thermal conduction
and toroidal magnetic field effect. We suppose a moderately
massive central black holes $\sim 10^{5-6} M_{\odot}$ under
critical accretion rate $1 M_{\odot} yr^{-1}$ (e.g.,Minishige et
al. 2000).

 This paper is organized as follows: Section 2, we present the equations of
magnetohydrodynamics as the basic equations and assumptions.
Self-similar solutions are presented in section 3. In section 4,
the continuum spectra are calculated and we regard
general properties of supercritical discs  and ultimately we show
the results in section 5.

\section{The Basic Equations}

For constructing the model, we used a set of coupled differential equations describing the law of conservation for a steady state, axi-symmetric ($\frac{\partial}{\partial \phi}=\frac{\partial}{\partial t}=0$) supercritical accretion disc. In cylindrical
coordinates $(r,\varphi,z)$, we used vertically integrate the flow equations, also, we suppose that all variables are only a
function of $r$. We ignore the relativistic effect and we use
Newtonian gravity. We adopt $\alpha$-prescription for viscosity
of accreting flow. The magnetic field was considered with toroidal
configurations.

The MHD equations are as the same as (Akizuki \& fukue 2006, Abbassi et al. 2008, GKA12):
\begin{equation}
\frac{1}{r}\frac{\partial}{\partial r}(r\Sigma V_r )=2 \dot{\rho}H
\end{equation}
\begin{equation}
V_r\frac{\partial V_r}{\partial r}=\frac{V_\varphi^{2}}{r}-\frac{G
M_{\ast}}{r^{2}}-\frac{1}{\Sigma}\frac{d}{dr}(\Sigma
c_s^{2})-\frac{c_A^{2}}{r}-\frac{1}{2\Sigma}\frac{d}{dr}(\Sigma
c_A^{2})
\end{equation}
\begin{equation}
r\Sigma V_r
\frac{d}{dr}(rV_\varphi)=\frac{d}{dr}(\frac{r^{3}\alpha c_{s}^{2}
\Sigma}{\Omega_{k}}\frac{d\Omega}{dr})
\end{equation}
\begin{equation}
\frac{GM}{r^{3}}H^{2}=c_s^{2}[1+\frac{1}{2}(\frac{c_A}{c_s})^{2}]=(1+\beta)c_s^{2}
\end{equation}
\begin{displaymath}
\frac{\Sigma V_r}{\gamma-1}\frac{dc_s^{2}}{dr}-2H V_r
c_s^{2}\frac{d\rho}{dr}=\frac{f\alpha\Sigma
c_s^{2}}{\Omega_k}r^{2}(\frac{d\Omega}{dr})^{2}-\frac{2H}{r}\frac{d}{dr}(r^{2}F_s)
\end{displaymath}
\begin{equation}
\frac{d}{dr}(V_r B_\varphi)=\dot{B_\varphi}
\end{equation}
where $V_{r}$ is the accretion velocity ($V_{r}<0$) and
$\Sigma=2\rho H$ is the surface density at a cylindrical radius
$r$, $\dot{\rho}$ is the mass loss rate per unit volume, H is the
disc half-thickness. The azimuthal equation of motion was integreated over $z$.
$\alpha$ is the viscous parameter,
$\Omega(=\frac{V_{\varphi}}{r})$ and $\Omega_{k}$ are the angular velocity and
the Keplerian angular speed respectively.

In this equation we have introduced
$\beta=\frac{P_{mag}}{P_{gas}}=\frac{1}{2}(\frac{c_A}{c_s})^{2}$
which is  magnetic field pressure over the gas pressure which
indicates the importance of magnetic field in the dynamics of
accretion flow and it is a free parameter in our model.
Following (GKA12) radiative pressure is ignored but it should take into account 
in the case of slim disc to have more realistic picture. As the same
as Akizuki \& Fukue 2006, Abbassi et al 2008 and GKA12 we can
chose two cases: Case 1: when the pressure is assumed to be the
gas pressure (thermal pressure). Case 2: when the pressure is
assumed to be the magnetic pressure plus the gas pressure. So in
the case 2 we replace $\alpha$ whit $\alpha(1+\beta)$ in all of
the equations. For simplicity in this investigation we use the
case 1.

We will follow how will change is not here the dynamical
and observational properties of the disc will change when the
$\beta$ varies is a physically reasonable interval. In the energy
equation we have cooling and heating term in the disc. We assume
the generated energy due to viscous dissipation plus the heat
conducted into the volume are balanced by the advection cooling.
The second term on right hand side of the energy equation
represents energy transfer due to the thermal conduction where
$F_s=5\Phi_s \rho c_s^{3}$ is the saturated conduction flux
(Cowie \& Makee 1977). Dimensionless coefficient $\Phi_s$ is less
than unity.

The induction equation (the final one)  represents the
interaction of fluid and magnetic field. Where
$\dot{B_\varphi}$ is the field scaping/creating rate due to
magnetic instability or dynamo effect.

\begin{figure}[h]
  \centering  
    \includegraphics[width=8cm]{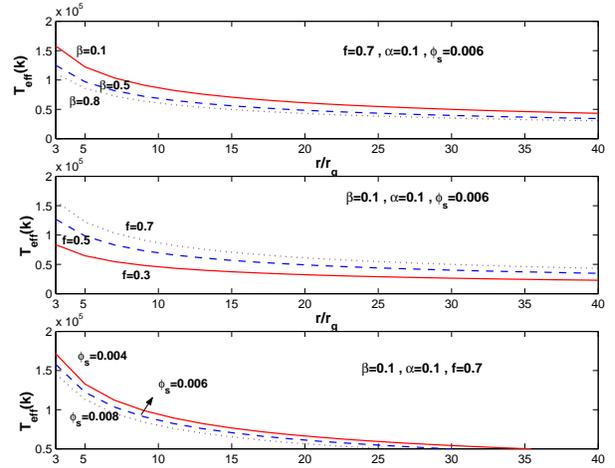}
       \caption{The surface temperature of LMC X-3 as a function of  dimensionless radius $(\frac{r}{r_{g}})$ for several values of (top panel:magnetic field ($\beta$) , middle panel:advction parameter ($f$),bottom panel:thermal conduction ($\phi_{s}$)).}
       \label{hallff}
\end{figure}

\begin{figure}[h]
  \centering  
    \includegraphics[width=8cm]{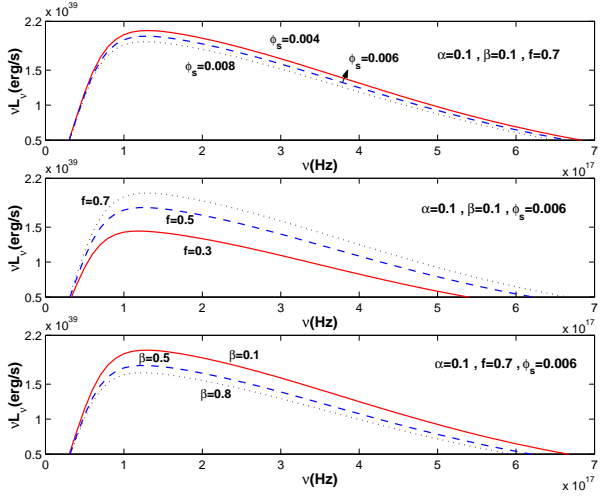}
       \caption{Continuum spectra of LMC X-3 for several values of (top panel:thermal conduction ($\phi_{s}$) , middle panel:advction parameter ($f$),bottom panel:magnetic field ($\beta$)).}
       \label{hallff}
\end{figure}

\begin{figure}[h]
  \centering  
    \includegraphics[width=8cm]{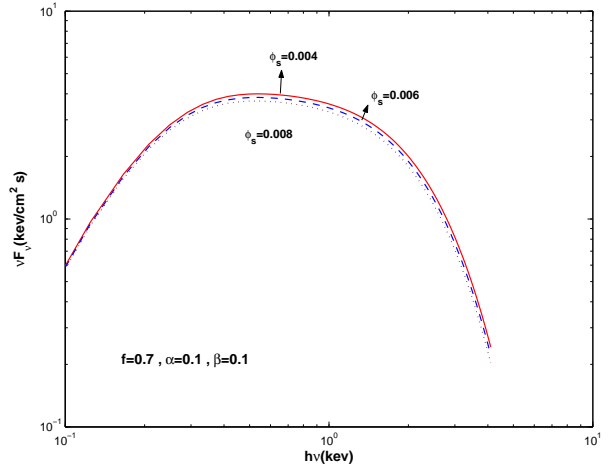}
       \caption{Observed flux of LMC X-3 with an inclination angles, $i=66$ for $\phi_{s}=0.004$ (solid lines), $\phi_{s}=0.006$ (dashed lines) and $\phi_{s}=0.008$ (dotted lines)}
       \label{hallff}
\end{figure}

\begin{figure}[h]
  \centering  
    \includegraphics[width=8cm]{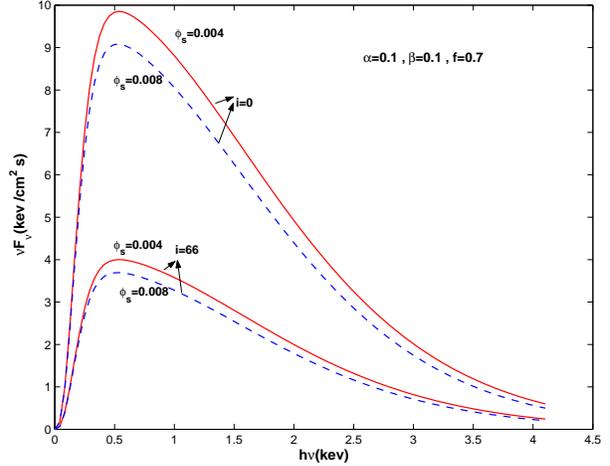}
       \caption{Observed flux of LMC X-3 with two inclination angles ($i=0$,$i=66$), $\phi_{s}=0.004$ (solid lines), $\phi_{s}=0.008$ (dashed lines)}
       \label{hallff}
\end{figure}

\section{Self-Similar Solutions}

To solve the MHD equations which they are introduced in last section and have physical
interpretation of a hot accretion flow under effect of thermal
conduction and toroidal magnetic field, we seek self-similar
solutions of the above MHD equations. The self-similar method is
familiar from its wide applications in astrophysics and it's
abilities to solve the full set of MHD equations. Usefulness of
the self-similar method is simplifying multi-dimensional problems
of geometric and kinematic complexity, making them analytically
tractable. The self-similar method is not able to reproduce the
exact solution of accretion flows, because no boundary conditions
have been taken into account. However, as long as we are not
interested in the behaviour of the flow near the boundaries, such
solutions are very useful. Following GKA12 solutions we have
introduce the velocities as follows

\begin{equation}
V_r(r)=-c_1 \alpha V_k(r)
\end{equation}
\begin{equation}
V_\varphi(r)=c_2 V_k(r)
\end{equation}
\begin{equation}
c_s^{2}=c_3 V_k^{2}
\end{equation}
\begin{equation}
c_A^{2}\frac{B_\varphi^{2}}{4\pi\rho}=2\beta c_3 \frac{GM}{r}
\end{equation}
where
\begin{equation}
V_k(r)=\sqrt{\frac{GM}{r}}
\end{equation}
and constant $c_1$, $c_2$ and $c_3$ are determined later from the
magnetohydrodynamic equations. From Eq. 4 we will obtain the disc half-thickness H as:
\begin{equation}
\frac{H}{r}=\sqrt{c_3 (1+\beta)}=\tan\sigma
\end{equation}
If we assume a power law form for the surface density $\Sigma$ as:
\begin{equation}
\Sigma=\Sigma_0 r^{s}
\end{equation}
we will obtain:

\begin{equation}
\dot{\rho}=\dot{\rho_0} r^{s-\frac{5}{2}}
\end{equation}
\begin{equation}
\dot{B_\varphi}=\dot{B_0} r^{\frac{s-5}{2}}
\end{equation}

Substituting the above self-similar transformation in to the MHD
equations of the system, we 'll obtain the following system of
dimensionless equations, which should be solve to having $c_1$,
$c_2$ and $c_3$: In case 1:
\begin{equation}
\dot{\rho_0}=-(s+\frac{1}{2})\frac{c_1
\alpha\Sigma_0)}{2}\sqrt{\frac{GM_\ast}{(1+\beta)C_3}}
\end{equation}
\begin{equation}
H=\sqrt{(1+\beta)c_3}r
\end{equation}
\begin{equation}
-\frac{1}{2}c_1^{2}\alpha^{2}=c_2^{2}-1-[s-1+\beta(s+1)]c_3
\end{equation}
\begin{equation}
c_1=3(s+1)c_3
\end{equation}
\begin{equation}
(\frac{1}{\gamma-1}-\frac{1}{2})c_1 c_3 =\frac{9}{4}f c_3
c_2^{2}-\frac{5 \Phi_s}{\alpha}(s-\frac{3}{2})c_3^{\frac{3}{2}}
\end{equation}

After algebraic manipulations, we obtain a forth order algebraic
equation for $c_1$:
\begin{equation}
D^{2}c_1^{4}+2DBc_1^{3}+(B^{2}-2D)c_1^{2}-(2B+A^{2})c_1+1=0
\end{equation}
where
\begin{equation}
D=\frac{1}{2}\alpha^{2}
\end{equation}
\begin{equation}
B={\frac{4}{9f}(\frac{1}{\gamma-1}-\frac{1}{2})-[s-1+\beta(s+1)][\frac{1}{3(s+1)}]}
\end{equation}
\begin{equation}
A=\frac{20\Phi_s}{9f\alpha}(s-\frac{3}{2})[\frac{1}{3(s+1)}]^{\frac{1}{2}}
\end{equation}

For the case $s=-\frac{1}{2}$ we have $\dot{\rho_0}=0$ which it
is correspond to no mass loss or wind in the hot magnetized flow,
Abbassi et al. (2008). In this work we focus on constant
accretion rate ( no outflow ) case but for $s>-\frac{1}{2}$ there
is wind or outflow and accretion rate is not constant. The
observation evidence shows that the outflow can exist in ADAF
(like $sgr A^{\star}$) and Slim disc ( like Seyfert 1 sources).
The properties of outflow examined by Xie \& Yuan (2008), Yuan,
Wu, Bu 2012, Yuan, Bu, Wu 2012 for ADAF, and Ohsuga, Mineshige,
et al. 2009 for ADAF, SSD and Slim discs . The outflow will
affect the dynamics, and the radiative output of the accretion
disc ( Xie \& Yuan 2012). It is clear that $c_1$ which is
determine the behaviour of the radial flow is specified by the
input parameter of the fluid such as $\alpha$, $\Phi_s$, $\beta$
and $f$. If we $c_1$ achieve from the above equation then we can
easily have $c_2$ and $c_2$ by:
\begin{displaymath}
c_2^{2}=\frac{4 c_1}{9f}[\frac{1}{\gamma-1}-\frac{1}{2}]
\end{displaymath}
\begin{equation}
+\frac{20\Phi_s}{9
f\alpha}(s-\frac{3}{2})[\frac{1}{3(s+1)}]^{\frac{1}{2}}
c_1^{\frac{1}{2}}
\end{equation}
\begin{equation}
c_3=c_1(\frac{1}{3(s+1)})
\end{equation}

GKA12 by adding one more constrain on solution presented by
Abbassi et al 2008, have shown the this new physical
condition places a constraint on the physically valid parametric
space. Their calculations are carried out in a range of
viscosities, thermal conduction, and magnetic field parameters
which they have acceptable physical behaviour. GKA12 have presented
dynamical behaviour of such flows by plotting $c_1$, $c_2$,
$c_3$. In this paper we want to further explore this system to
investigate the possible role of thermal conduction and magnetic
field on the observational appearance of hot magnetized flow. In
the next section we will follow this idea.

\section{radiation properties}
We have a complete set of equations which describe the global dynamics of hot magnetized flow.
Using the results obtained in the previous section we will able reproduce observational appearance of the hot magnetized flow.
The surface flux and disc luminosity of the discs are derived as
follow: By assuming a dominance of the radiation pressure, we can
write the height-integrated pressure $\Pi(=\Sigma c_{s^{2}})$ and
the averaged flux F as:
\begin{equation}
\Pi=\Pi_{rad}=\frac{1}{3}a T_{c}^{4}2H=\frac{8H}{3c}\sigma
T_{c}^{4}
\end{equation}

\begin{equation}
F=\sigma
T_{c}^{4}=\frac{3c}{8H}\Pi=\frac{3}{8}c\Sigma_{0}\sqrt{\frac{c_{3}}{1+\beta}}G
M r^{s-2},
\end{equation}

where $\sigma$ is the Stefan-Boltzman constant. The optical
thickness of the disc as the same as height-integrated pressure
and averaged flux has a radial dependency as :

\begin{equation}
\tau=\frac{1}{2}k \Sigma=\frac{1}{2}k \Sigma_{0}r^{s}
\end{equation}
where $k (=0.4 cm^{2}g^{-1})$ and
$\tau=\frac{16\sqrt{6}}{\alpha}\sqrt{\frac{r}{r_{g}}}$ are the
electron-scattering opacity and Schwarzschild respectively.

Hence, the effective flux and the effective temperature of the
disc surface become:
\begin{equation}
\sigma T_{eff}^{4}=-\frac{16 \sigma T^{4}}{3 K
\rho}\frac{\partial T}{\partial z}\approx \frac{4 \sigma}{3\tau}
T^{4}
\end{equation}
\begin{equation}
\sigma T_{eff}^{4}=\frac{\sigma
T_{c}^{4}}{\tau}=\frac{3c}{4k}\sqrt{\frac{c_{3}}{1+\beta}}\frac{G
M}{r^{2}}=\frac{3}{4}\sqrt{\frac{c_{3}}{1+\beta}}\frac{L_{E}}{4\pi
r^{2}}
\end{equation}

\begin{equation}
T_{eff}=(\frac{3L_{E}}{16\pi
\sigma}\sqrt{\frac{c_{3}}{1+\beta}})^{\frac{1}{4}}r^{\frac{-1}{2}}
\end{equation}
where $L_{E}=4\pi c \frac{G M}{k}$ is the Eddington luminosity.

By integrating these equations radially we will have the disc
luminosity as:
\begin{equation}
L_{disc}=\frac{3}{4}\sqrt{\frac{c_{3}}{1+\beta}}L_{E}\ln\frac{r_{out}}{r_{in}}
\end{equation}
where $r_{in}\sim 3 r_{g}$ , $r_{out}=2\times 10^{5} r_{g}$
(Fukue, 2000). This equation represents, the disc's
luminosity which is affected by magnetic field explicitly through
the $\beta$, but it would be affected by thermal conduction,
advection parameter and viscosity through the $c_{3}$ implicitly.

The simplest spectral model we can construct for an accretion
disc if we assume the inflowing matter in a thermal equilibrium.
We have assumed the accretion flow radiates away locally like a
black-body radiation. We expect that the hottest regions at small
$r$ to radiate at the highest energies. This simple spectral
model is referred to as the multi-colour disc. For the present
purpose, we assume that the disc surface radiates
black-body radiation $B_{\nu}$ with temperature
$T_{eff}(r)$. Then we can equate the disk radiative flux
$F(r)$ to a black body flux $\sigma T_{eff}^{4}(r)$ where
$T_{eff}(r)$ is the surface temperature. Since $\sigma
T_{eff}^{4}(r)=\int_{0}^{\infty}\pi B_{\nu}(T_{eff}(r)) d\nu$, We
can equate radiative flux $F_{\nu}(r)=\pi B_{\nu}(T_{eff}(r))$.
Then, the continuum spectrum (luminosity per frequency) $L_{\nu}$
can be calculated (see , e.g., Kato et al. 1998) by:
\begin{equation}
L_{\nu}=2\int_{r_{in}}^{r_{cr}}\pi B_{\nu}(r) 2\pi r dr
\end{equation}
where a factor 2 comes from that the radiation will radiated from
both side of the disc. We will integereated luminocity in a
resonable interval: $(r_{cr}=100-1000 r_{g})$. We have used
black-body function:

\begin{equation}
B_{\nu}(r)=\frac{2h}{c^{2}}\frac{\nu^{3}}{e^{\frac{h\nu}{k_{B}T_{eff}(r)}}-1}
\end{equation}

The disc is believed to be fundamentally antisymmetric. When
viewed at an angle other than face-on (from above or below the
plane exactly), the circular disc appears elliptical. The greater
the ellipticity, the more the galaxy's disc in inclined with
respect to our line of sight. Here we are studding of disc viewed
at different inclinations. Observed flux from an
accretion disc depends on the distance $D(=48.1 kpc)$
and it's inclination angle $i$ as:
\begin{equation}
F^{obs}_{\nu}=\frac{L^{obs}}{\pi D^{2}}=\frac{\cos i
\frac{L_{\nu}}{2}}{\pi D^{2}}
\end{equation}
\textbf{This equation can be used for standard discs. }

\begin{figure}[h]
  \centering  
    \includegraphics[width=8cm]{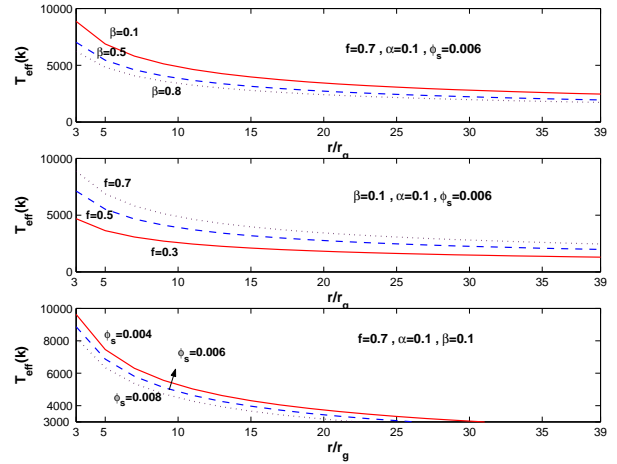}
       \caption{The surface temperature of narrow-line seyfert 1 galaxies as a function of  dimensionless radius $(\frac{r}{r_{g}})$ for several values of (top panel:magnetic field ($\beta$) , middle panel:advction parameter ($f$),bottom panel:thermal conduction ($\phi_{s}$)).}
       \label{hallff}
\end{figure}

\begin{figure}[h]
  \centering  
    \includegraphics[width=8cm]{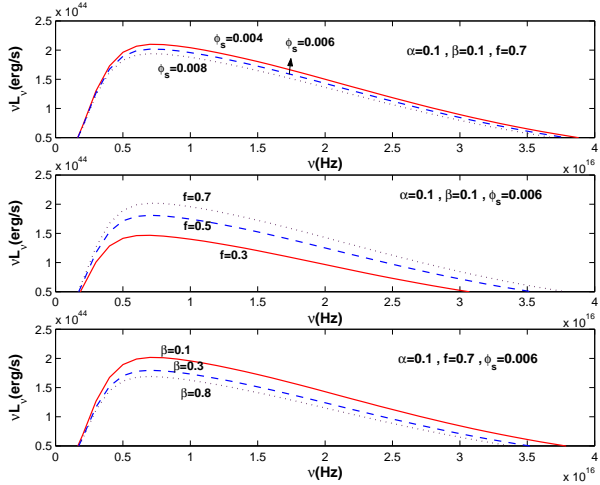}
       \caption{Continuum spectra of narrow-line seyfert 1 galaxies for several values of (top panel:thermal conduction ($\phi_{s}$) , middle panel:advction parameter ($f$),bottom panel:magnetic field ($\beta$)).}
       \label{hallff}
\end{figure}

\begin{figure}[h]
  \centering  
    \includegraphics[width=8cm]{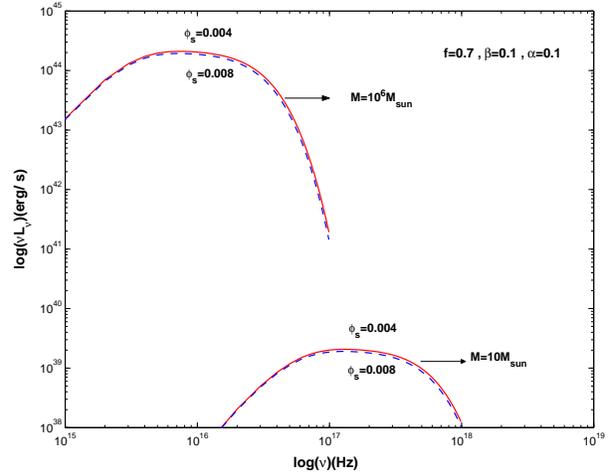}
       \caption{Continuum spectra of narrow-line seyfert 1 galaxies with $m=10^{6}m_{\odot}$ and LMC X-3 with $m=10 m_{\odot}$ for the two thermal conduction parameter $\phi_{s}=0.004$ , $\phi_{s}=0.008$}
       \label{hallff}
\end{figure}

\section{Results}
In this section we apply the present model of hot magnetized accretion
disc to two astronomical objects, including an accretion disc
around a black hole, which are supposed to be under critical
accretion: LMC X-3 and narrow-line seyfert 1 galaxies.

LMC X-3 is a powerful source of X-rays located in the Large Magellanic Cloud (LMC). This X-ray source is associated with a binary system with an orbital period of 1.7 days (close binary). The visible component is a main sequence B3 star whose shape has been severely distorted by the gravitational field of its companion. Although not unambiguous, the mass of the compact object is estimated to be order of 10 solar masses and more likely is considerably higher, making this one of the best black hole candidates. We used LMC X-3 values to test our model.

In figure 1 the surface temperature ($T_{eff}$) is plotted as a
function of the dimensionless radius ($\frac{r}{r_{g}}$) for LMC
X-3. It is obvious that the surface temperature decreases as
$\frac{r}{r_{g}}$ increases. The top panel shows the effect of
magnetic field on the surface temperature. We see that the surface
temperature decreases by adding magnetic field parameter $\beta$.
In the middle panel we show that surface temperature increases by
increasing advection parameter. And finally in the lower panel we have plotted
surface temprature as a function of radial distance for several values of
thermal conduction parameter, $\phi_s$. Slope of the temperature gradient decreases
when we add the role of thermal conduction, by adding $\phi_s$, as expected. This is
acceptable because the increasing of thermal conduction causes
that the more energy transfers to outer of disc and so reduce the temperature gradient.
To having this panel we just used physically allowed parameter space which was introduced by GKA12.

The radiation spectrum of the hot magnetized flow is shown in
figure 2 for several values of thermal conduction, advection
parameter and magnetic field from top to bottom. The central
black-hole mass is fixed as $10 M_{\odot}$, while the mass
accretion-rate $\dot{M}$ is equal $0.6 \dot{M}_{Edd}$. As can be
seen in figure 2, the maximum of $\nu L_{\nu}$ is always of the
order of the Eddington luminosity ($L_{E}=10^{39}erg s^{-1}$).
This agrees with Fukue (2004) and observed value of LMC X-3. As we
see by increasing of thermal conduction, the surface temperature
decreases and therefore luminosity of disc also decreases. But
advection parameter acts oppose to thermal conduction. We
expect that by decreasing the advection parameter, $f$, more energy
radiated away so the total luminosity will be higher while
we see the opposite result in figure 2 and 6, seems this behavior comes from the lack of radiation pressure. So this results indicate that in ordetr to have a realistic picture we should
add the radiation pressure in the dynamics of hot magnetized flow. 

We plotted the observed flux of LMC X-3 for inclination angle $i=66^{\circ}$ for
different values of thermal conduction in figure 3.
In figure 3,4 we evaluate the effect of thermal conduction on the
observed flux. Figure 4 we have shown that by increasing the
inclination angle $i$ for two thermal conduction parameters, the radiation spectrum of the disc $\nu
F_{\nu}$ will increase. In both of two inclination angles we can
easily see by increasing of thermal conduction $\phi_{s}$, the
observed flux decreases. In this panel we have the same result
with figure 2. The classical soft-state spectrum of the X-ray
binary LMC X-3 in dominated by a $\sim 0.5 kev $. This result
agrees with Straub et al. (2011). Davis et al.
(2006), have plotted the spectrum of LMC X-3  for $i=60^{\circ}$,
$D=52 kpc $, and they have shown that the observed flux is approximately $1
\frac{kev}{cm^{2} s}$ and the peak of the spectrum is located in
$E=1 kev$.

The same panels are shown for narrow-line seyfert 1
galaxies in figure 5 ,6. Seyfert galaxies are characterized by
extremely bright nuclei, and spectra which have very bright
emission lines of hydrogen, helium, nitrogen, and oxygen. These
X-ray emission may come from the surface of the accretion disc
itself. The central black hole mass is $10^{6}M_{\odot}$ and
$\dot{M}\sim 10^{3} \dot{M}_{Edd}$. In figure 5 as can be seen
for this object temperature is lower than LMC X-3. This result is
according to the result of Abramowicz et al.(1988). We show in
figure 6 the maximum of $\nu L_{\nu}$ is always of the order of
the Eddington luminosity ($L_{E}=10^{44}erg s^{-1}$) and disc
spectrum is peaking up $\nu=0.5 \times 10^{16} Hz$. These results
are agree with Fukue (2004).

In figure 7 we compared the solution for these two
objects. We have shown that for the same thermal conduction
parameter when the mass of BH decreases, the spectrum is setting
in high frequency and high energy while the luminosity decreases
and we have hard-low state.

\section{SUMMERY AND CONCLUSION}

We have presented a simple study showing that self-similar
solution of a hot magnetized flow under the effect of thermal
conduction. Purely toroidal magnetic field configuration is
assumed. It was assumed the disc is axially symmetric and static
with the $\alpha$-prescription of viscosity. The relativistic
effects and self-gravity of the disc are ignored. The
weak collisional nature of a hot accretion flow was confirmed by
(Tanaka \& Menou 2006; Abbassi et al. 2008), so we have adopted a
saturated form of thermal conduction as a possible mechanism for
energy transportation. Observational appearance of such disc was
investigated with a physically limited parameter space presented
by GKA12.

GKA12 imposed condition puts some physical constraint on
self-similar solutions presented by Abbassi et al 2008. As a
result, in order to have a physically valid solution, only a small
part of the parameter space should be considered. After assuring
the validity of our solutions by using the proper values of input
parameters $\alpha$, $\beta$, $\phi_s$ and
f we have investigated the influence of thermal conduction on the
observational appearance of hot magnetized accretion flows.

It is shown that the temperature gradient will decrease when
thermal conduction plays an important role.
increases when thermal conduction plays an important
role. Radiation spectrum of the hot magnetized flow was plotted for
two real cases: narrow-line seyfert 1 galaxies and X-ray
binary LMC X-3. The theoretical solution is compatible with
observed x-ray spectra of these two cases.

So far we have considered only the most basic of hot magnetized
flow: one-fluid, axially symmetric, and steady. The real hot
magnetized flow is a much richer medium, providing numerous
examples of important astrophysical phenomena on all scales.
Thus, if we assume a two-temperature structure for the ions and
electrons in the discs, then we may expect that the ions
and electron temperatures decouple in the inner regions where is
so hot, which will modify the role of conduction consequently will
modify the dynamical and observational appearance of the disc.
One of the other limitation of this solution is the anisotropic
character of conduction in the presence of a magnetic field which
is indicated by Balbus (2001).

\end{document}